# Link Prediction in Multiplex Networks based on Interlayer Similarity

Shaghayegh Najari[a], Mostafa Salehi[a,b] *, Vahid Ranjbar[a], Mahdi Jalili[c]

[a] Faculty of New Sciences and Technologies, University of Tehran, Tehran, Iran
[b] School of Computer Science, Institute for Research in Fundamental Science, Tehran, Iran
[c] School of Engineering, RMIT University, Melborn, Australia

## Highlights

- We propose an interlayer similarities measure for multiplex networks.
- We introduce a link prediction framework based on interlayer similarities for multiplex networks.
- We examine the impact of interlayer similarities on link prediction in synthetic and real multiplex networks.

## Abstract

Some networked systems can be better modelled by multilayer structure where the individual nodes develop relationships in multiple layers. Multilayer networks with similar nodes across layers are also known as multiplex networks. This manuscript proposes a novel framework for predicting forthcoming or missing links in multiplex networks. The link prediction problem in multiplex networks is how to predict links in one of the layers, taking into account the structural information of other layers. The proposed link prediction framework is based on interlayer similarity and proximity-based features extracted from the layer for which the link prediction is considered. To this end, commonly used proximity-based features such as Adamic-Adar and Jaccard Coefficient are considered. These features that have been originally proposed to predict missing links in monolayer networks, do not require learning, and thus are simple to compute. The proposed method introduces a systematic approach to take into account interlayer similarity for the link prediction purpose. Experimental results on both synthetic and real multiplex networks reveal the effectiveness of the proposed method and show its superior performance than state-of-the-art algorithms proposed for the link prediction problem in multiplex networks.

*Keywords:* Link Prediction, Multiplex network, Multirelational Network, Machine Learning, Complex Network.

## 1. Introduction

Complex networks [1] are powerful tools to represent complex relationships between individual units [2]. Network analysis is an emerging branch of science and many natural and human-made complex systems can be modeled as networks where a number of individual units are connected through interaction links [3,4]. Example of such systems include online social networks, the Internet, World Wide Web, the human brain, power grids, transportation and water distribution networks. One of the heavily studied topics in network analysis and mining is link prediction problem, that is to predict forthcoming or missing links of a network [5], which has many potential applications in bioinformatics and medicine [6, 7], social network [5] and e-commerce [8-11]. In online social networks, for instance, one can use the link prediction algorithms to suggest friendship links to individuals [12]. Biological networks are another example where the link prediction algorithms can be used to discover missing links between proteins to avoid further experiments, which can be costly and time-consuming [13].

\* *Corresponding author: mostafa salehi at Faculty of New Sciences and Technologies, University of Tehran, Tehran, Iran, E-mail: mostafa_salehi@ut.ac.ir.*
E-mail addresses: najari.shaghayegh@ut.ac.ir, mostafa_salehi@ut.ac.ir, vranjbar@ut.ac.ir, mahdi.jalili@rmit.edu.au

Various approaches have been introduced to predict links in networked structures, which can be generally categorized into two classes: similarity-based and learning-based [12]. Similarity-based methods do not use any form of learning, and each pair of non-connected nodes takes a similarity score where the top-ranked relationships are more likely to appear in the future. Learning-based methods on the other hand use statistical machine learning techniques, often resulting in better prediction performance than similarity-based methods but with the price of more computational complexity [14]. In statistical methods, the likelihood of an edge between two non-adjacent nodes is obtained through local or global nodal measures. In machine learning based methods however the problem is solved as a classification problem where there are two classes, one for the existence of edges and another one for non-existence. To this end, a number of features are first extracted, and then a learning algorithm is used to solve the classification problem.

Traditionally, complex systems are modeled as monolayer networks, for which nodes and edges are all from the same type. However, many real systems might develop their connections in multiple layers [15, 16]. Individuals might have friendship links in different social media platforms, where links in each platform can be considered in a separate layer. Cities can be connected through air, road and rail networks, each of which can be considered as a separate layer [17]. Multilayer networks, also known as multiplex, heterogeneous networks or network of networks is the new way of modeling such complex systems with multiple layers of interactions between units. Real multilayer networks exhibit strong correlation between the nodes features in different layers [15,18]. This indicates that, neglecting the influence of multiple layers on predicting links in one of layers might lead to loss of significant information. Hristova at et al. [19] considered a two-layer network of a number of users of Twitter and Foursquare multiplex network, and showed that considering the information of the layers can significantly improve the performance of the link prediction problem as compared to the cases where the intra-layer links are predicted only based on the information in the parent layer.

Recently, a number of methods have been proposed to solve the link prediction problem in multilayer and multiplex networks. Davis at et al. [20] proposed a probabilistically weighted extension of the Adamic-Adar [21] measure in a supervised learning manner for the link prediction in multilayer networks, and showed that the supervised learning methods work much better than the unsupervised ones. Yang et al. [22] proposed a probabilistic method, called Multi Relational Influence Propagation (MRIP), to predict links in heterogeneous networks for which the nodes and edges can be of different types. Jalili et al. [23] introduced a method that considers features extracted from meta-paths and used machine learning algorithms to solve the problem. In another work, Pujari et al. [24] used entropy-based features that used both interlayer and intralayer information. In this paper, we propose a framework to systematically consider interlayer similarity for the link prediction problem in multilayer networks. Furthermore, we propose a new framework based on interlayer similarities in multiplex networks. Our experimental results show that the proposed method outperforms state-of-the-art link prediction methods.

2. **Preliminaries**

As discussed before, ignoring interlayer information or merging them with intralayer feature might lead to loss of some information in a multiplex network [26]. A number of methods have been proposed in the literature to consider information across layer for the link prediction task [19, 20, 22-24]. However, they have not considered interlayer similarity, and our aim in this work is to propose a methodology to systematically take into account the interlayer information for the link prediction task. To better understand the problem, let us first introduce the notions used for multiplex networks and formally define the link prediction problem in them.

Let us denote multiplex networks as $G(L_1 L_2 \dots L_N)$, where $L_i = L(V, E_i)$ represents a layer from multiplex network, in which $V$ is a set of nodes (the same across the layers) and $E_i (i = 1, 2, \dots N)$ represents the set of links of type $i$. $N$ is the number of layers. The link prediction problem in a multiplex network is to predict the potential links between each node pairs in any of the layers by using multiplex network features extracted from intralayer and interlayer relationships (Fig. 1). A natural approach for the link prediction problem in multiplex networks is to simply extend the methods originally proposed for monolayer networks. In the following, we provide a brief explanation some of features that have been commonly used by various link prediction works. We also discuss a number of metrics that can be used to measure interlayer similarity in multiplex networks.

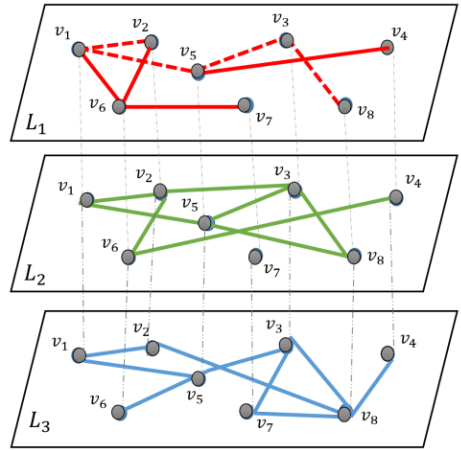

Fig 1. Network **G** with 3 layers $\mathbf{L_1, L_2, L_3}$. The link prediction problem in this network is to predict the links in one of layers by using features extracted from intralayer and interlayer relationships.

### 2.1 Features for the link prediction in monolayer networks

Various local and global structural features have been proposed for the link prediction task in monolayer networks. These features are properties of the end nodes topping to edges, returning the existence probability for the links. These features have been originally proposed for monolayer networks. Let's call these features as "intralayer features", which are generally categorized into neighborhood-based and path-based features.

### 2.1.1 Neighborhood-based metrics

Many link prediction methods have been proposed based on various node neighborhood measures. These features are based on local nodal properties as they use only information available from local proximity of the nodes. Common Neighbors (CN) of two nodes is the number of common neighbors between them, i.e. the set of nodes that are connected to either of these nodes and their path length is 2 [25]. Jaccard Coefficient (JC) is the number of common neighbors divided by the total number of neighbors of each pair [25]. Another variation of common neighbors is the Adamic-Adar (AA) measure that is to inversely weight the common neighbors with respect to the degree of the nodes [21]. The Preferential Attachment (PA) is another simple method for scoring each pair of nodes, which is the product of their degrees [27]. Resource Allocation (RA) and Local Path (LP) [28] are two well-known metrics that are compared with the proposed method. All of these measures are based on local properties of nodes.

*2.1.2  Path-based metrics*

These features are based on paths between nodes which require global information of the networks. Katz measure [29] is based on shortest paths and directly counts the number of paths between the nodes penalizing longer paths. Another measure in this category is Page Rank (PR) [30] that represents the significance of a node in the network based on the significance of its incoming neighbors. Rooted Page Rank (RPR) [5] is another metric used in the link prediction tasks, which has been derived originally from Page Rank. RPR corresponds to the visiting a specific node during a random walk from the source.

*2.2  Interlayer Similarity Measures*

In this work, we aim to use interlayer similarity systematically to improve the link prediction performance in multiplex networks. There are a number of approaches to calculate the similarity between layers of multiplex networks [31-34], from which we consider the commonly used ones. We also proposed a new interlayer similarity metric and show that it can effectively improve the link prediction performance. For the same of simplicity, we will define some of these metrics for a two-layer network. Let $G_T(L_1, L_2)$ be a multiplex network containing $N$ nodes with two layers $L_1$ and $L_2$.

- Degree- degree Correlation (DDC)

The nodes may have different degrees across the layers. Degree-degree Correlation (DDC) captures interlayer correlation of degrees across different layers, as follows [31]:

$$DDC = \frac{\sum_{k_{L_1}} \sum_{k_{L_2}} \left( k_{L_1} k_{L_2} \left( p(k_{L_1}, k_{L_2}) - \left( \sum_{k_{L_1}} p(k_{L_1}, k_{L_2}) \right) \left( \sum_{k_{L_2}} p(k_{L_1}, k_{L_2}) \right) \right) \right)}{\sum_{k_{L_2}} k_{L_2}^2 \sum_{k_{L_1}} p(k_{L_1}, k_{L_2}) - \left( \sum_{k_{L_2}} k_{L_2} \sum_{k_{L_1}} p(k_{L_1}, k_{L_2}) \right)^2} \quad (1)$$

where $p(k_{L_1}, k_{L_2})$ is the probability that a random node from has degree $k_{L_1}$ in layer $L_1$ and $k_{L_2}$ in layer $L_2$. If $DDC < 0$, two layers have negative correlation, while they have positive correlation if $DDC > 0$, and un-correlation if $DDC = 0$. This metric can be used as an interlayer similarity.

- Betweenness-based (BW)

The centrality of nodes indicates their importance and vitality in the network. Various centrality measures have been proposed in the literature to capture this. Betweenness centrality that has been used frequently to measure node vitality, is based on the number of times a node appears in the shortest paths between nodes [35]. In other words, the betweenness centrality for each node is the number of the shortest paths that pass through that node. Distance of betweenness centrality is defined as:

$$D_{BW_i} = |BW_i^{(L_1)} - BW_i^{(L_2)}| \quad (2)$$

where $BW_i^{(L_1)}$ is the betweenness centrality of node $i$ in $L_1$. One can define betweenness similarity of node $i$ across two layers as:

$$S_{BW_i} = 1 - D_{BW_i} \qquad (3)$$

By averaging ~~of~~ $S_{BW_i}$ over all nodes~~,~~ one can obtain average or normalized betweenness similarity as:

$$S_{BW} = \frac{\sum_{i=1}^{N} S_{BW_i}}{N} \qquad (4)$$

- Clustering Coefficient-based (CC)

Clustering coefficient is a metric to measure the extent to which nodes in a network tend to cluster together. Batisteun et al. [32] provided two new methods for calculating the clustering coefficient in multiplex networks. The clustering coefficient of node $i$ in each of layers is defined as:

$$CC_i^{(L)} = \frac{\sum_{j \neq i,\ m \neq i} a_{ij}^{(L)} a_{jm}^{(L)} a_{mi}^{(L)}}{\sum_{j \neq i,\ m \neq i} a_{ij}^{(L)} a_{mi}^{(L)}} \qquad (5)$$

where $CC_i^{(L)}$ is the clustering coefficient of node i in layer L, $a_{ij}$ present the status of link $(i,j)$ in a network, it is 1 if the link $(i,j)$ exists, or 0 otherwise. The distance between clustering coefficient of node $i$ across the two layers is:

$$D_{CC_i} = |CC_i^{(L_1)} - CC_i^{(L_2)}| \qquad (6)$$

Clustering similarity of node $i$ is:

$$S_{CC_i} = 1 - D_{CC_i} \qquad (7)$$

The average of the above over all nodes is the clustering similarity of the multiplex network, as following:

$$S_{CC} = \frac{\sum_{i=1}^{N} S_{CC_i}}{N} \qquad (8)$$

- Average Similarity of the Neighbors (ASN)

Each node in $G_T$ may be connected to three types of edges: intralayer links in $L_1$, those in $L_2$ and interlayer links between $L_1$ and $L_2$. For any node $i$, $k_{L_1}(i)$ is the total number of connections in $L_1$, $k_{L_2}(i)$ is the total number of connections in layer $L_2$ and $k_L(i)$ is the total number of common connection in $L_1$ and $L_2$. The average similarity of the neighbors in the multiplex network $G_T$ is defined as [31]:

$$ASN = \frac{\sum_i k_L(i)}{\sum_i \left(k_{L_1}(i) + k_{L_2}(i) - k_L(i)\right)} \qquad (9)$$

- Asymmetric ASN (AASN)

The average similarity of neighbors, as explained above, is a reciprocal relationship. A natural extension is to normalize this metric against the total number of links in the layers to account for network density. The intuition behind such a normalization is that by assuming high enough similarity between the layers, the denser are a layer, the more is the information it includes the link prediction of other layers. Following this intuition, we propose an Asymmetric ASN (ASNN), and define the interlayer similarity based on this concept as:

$$S_{AASN(L_1,L_2)} = \frac{\sum_i k_L(i)}{\sum_i k_{L_1}(i)}, \qquad (10)$$

$$S_{AASN(L_2,L_1)} = \frac{\sum_i k_L(i)}{\sum_i k_{L_2}(i)} \qquad (11)$$

where $S_{AASN(L_1,L_2)}$ is the interlayer similarity of layer $L_1$ with respect to layer $L_2$. Indeed, this metric is used for predicting links in $L_2$ taking into account the information available in $L_1$. Similarly, AASN-based interlayer similarity for node pair $(i,j)$ is defined as:

$$S_{AASN(L_1,L_2)}(i,j) = \frac{k_L(i) + k_L(j)}{k_{L_1}(i) + k_{L_1}(j)} \qquad (12)$$

$$S_{AASN(L_1,L_2)}(i,j) = \frac{k_L(i) + k_L(j)}{k_{L_2}(i) + k_{L_2}(j)} \qquad (13)$$

## 3. Link Prediction accounting Interlayer Similarity (LPIS) – The Proposed Framework

Fig. 2 represents the framework of the proposed link prediction method, which is based on using both intralayer features and interlayer similarity. In the proposed framework, the intralayer predictor calculates the probability of link existence by using intralayer features. Then these probabilities along with interlayer similarity are given to a synthesizer to calculate the final probability values for link existence. Algorithm1 shows a pseudo-code of the proposed link prediction procedure. We provide details of these steps in the following.

### 3.1 Intralayer Link Predictor

The intralayer link predictor uses only intralayer information for the link prediction. One can use classification or probabilistic methods as the intralayer link predictor in this framework. In classification based methods, the link prediction problem is considered as a classification problem with two classes. Then, a number of features are considered and a proper classifier such as Support Vector Machines (SVM), Naive Bays, K-Nearest Neighbors (KNN), or logistic regression is used to solve the problem. In probabilistic models, latent features are used to obtain a probabilistic model, resulting in the probability of link existence. A number of probabilistic methods have been proposed in the literature for the link prediction in monolayer networks. Examples include local relation models [36], hierarchical probabilistic models [37], and probabilistic relation models such as Bayesian relations [38] or Markova relation network [39]. Our proposed model in this work is a probabilistic-based model where intralayer features and interlayer similarity are used to obtain the final probabilities of link existence.

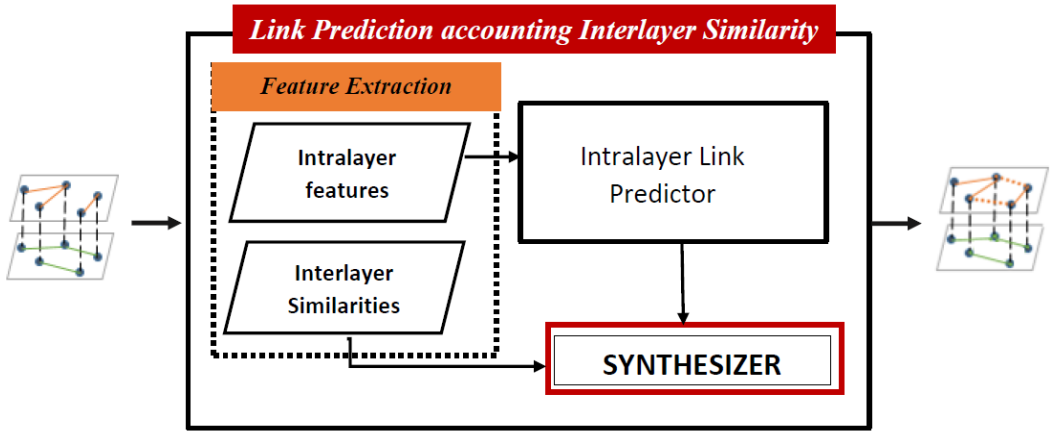

Fig 2. Proposed Framework for link prediction in multiplex networks.

---

**ALGORITHM 1:** Link Prediction accounting Interlayer Similarity (LPIS)

---
% $L_m$ and $L_k$: layers of multiplex network
% $(i,j)$: node pair
% $S_{L_m L_k}$: interlayer similarity between layer $L_m$ and $L_k$
$I_{L_k}$: Adjacency matrix of $L_k$
**for** $L_m$ in a multiplex-network:
 $p^{intra} \leftarrow$ Compute link existence probabilities by using intralayer features
**end**
**for** $L_m$ in a multiplex-network:
    **for** any $(i,j)$:
$$p_{L_m}^{inter}(i,j) = \begin{cases} \sum_{k \in \{1,2,\dots,n\},\ m \neq k} p_{L_k}^{intra}(i,j) \times S_{L_m L_k}(i,j), & if\ I_{L_k}(i,j) = 1 \\ \sum_{k \in \{1,2,\dots,n\},\ m \neq k} \left(1 - p_{L_k}^{intra}(i,j)\right) \times \left(1 - S_{L_m L_k}(i,j)\right), & if\ I_{L_k}(i,j) = 0 \end{cases}$$
    **end**
    **for** any $(i,j)$:
    Normalize $p_{L_m}^{inter}(i,j)$ as $(p_{L_m}^{inter}(i,j) / \max_{(i,j)} p_{L_m}^{inter})$
    $p_{L_m}^{total}(i,j) = (1-\alpha).p_{L_m}^{intra}(i,j) + \alpha.p_{L_m}^{inter}(i,j)$
    Labeled data (existence or non-existence of a link)←Labling (i, j) by thresholding
    **end**
**end**
**return** Labeled data

---

### 3.2 Synthesizer

Our main aim in this manuscript is to study whether considering interlayer information can boost the link prediction performance in multiplex networks. The proposed framework includes a synthesizer that combines intralayer and interlayer information to produce meaningful information for the link prediction task. Given a multiplex network with layers $\{L_1, L_2, \dots, L_n\}$, where $n$ is number of layers, the probability of link existence in layer $L_m (m \in \{1, 2, \dots, n\})$ is:

$$p_{L_m}^{total}(i,j) = (1-\alpha).p_{L_m}^{intra}(i,j) + \alpha.p_{L_m}^{inter}(i,j) \tag{14}$$

where $p_{L_m}^{total}(i,j)$ is the probability of existence of link $(i,j)$ in layer $L_m$, which is indeed the output of the synthesizer that combines the interlayer and intralayer features using a control parameter $\alpha$. Parameter $\alpha$ needs to be optimized for each task. In the above equation, $p_{L_m}^{intra}(i,j)$ is the existence probability of link $(i,j)$ in layer $L_m$ based on only of intralayer information, i.e. the output of the intralayer link predictor. $p_{L_m}^{inter}(i,j)$ is the existence probability of link $(i,j)$ in layer $L_m$, that is obtained based on interlayer information, as:

$$p_{L_m}^{inter}(i,j) = \begin{cases} \sum_{k \in \{1,2,\dots,n\},\ m \neq k} p_{L_k}^{intra}(i,j) \times S_{L_m L_k}(i,j), & if\ I_{L_k}(i,j) = 1 \\ \sum_{k \in \{1,2,\dots,n\},\ m \neq k} \left(1 - p_{L_k}^{intra}(i,j)\right) \times \left(1 - S_{L_m L_k}(i,j)\right), & if\ I_{L_k}(i,j) = 0 \end{cases} \tag{15}$$

where $S_{L_m L_k}(i,j)$ is the similarity of link $(i,j)$ between layers $L_m$ and $L_k$. When the network has more than two layers, the similarity of the link is obtained across all layers by obtaining the similarities for all layer pairs. In the above equation, $I_{L_k}$ is adjacency matrix of layer $L_k$, where $I_{L_k}(i,j) = 1$ if there is a link between nodes i and j in this layer, and $I_{L_k}(i,j) = 0$ otherwise. The intuition behind the above formulation is as follows. If link $(i,j)$ is present in layer $L_k$, $p_{L_m}^{inter}(i,j)$ is taken as the amount of the interlayer similarity between $L_m$ and $L_k$. This indicates that if link $(i,j)$ exists in $L_k$ and there is high similarity between this layer and $L_m$, it is likely that the link will also exist in layer $L_m$, and the higher is the similarity of the layers, the higher is the likelihood of the link existence in the other layer.

## 4. Experimental Results

We evaluate the proposed method on both synthetic and real datasets. In this work, we consider the link prediction problem as a classification task and use Logistic regression classifier with intralayer features as Adamic-Adar, Jaccard, preferential attachment indices, which has already been used for the link prediction task in monolayer networks [15].

Generally, the link prediction task is predicting the links are not present in the train data and may be created in the future network (test data). As mentioned, for predicting the links in each layer by using the LPIS framework, we need an intralayer link prediction model, which can be any desired intralayer link prediction model. The model used in this work is a classification model. We evaluate the performance of the predictions by Area Under the (Receive Operating Characteristics (ROC)) Curve (AUC) metric. The ROC curve is created by plotting the true positive rate against the false positive rate at various threshold settings. AUC is the probability that a classifier will rank a randomly chosen positive instance higher than a randomly chosen negative one. Generally, AUC is a prominent analyzer of binary classifiers and has been frequently used in the link prediction tasks.

*4.1 Synthetic Networks*

We first study the performance of the proposed framework in synthetic multiplex networks. To this end, we construct a two-layer network and apply the proposed link prediction methods. Given a multiplex network with two layers $L_1$ and $L_2$, we generate layer $L_1$ by Barabasi-Albert (BA) [34] or Watts-Strogatz (WS) [40] models. These two models have been frequently used in the literature to construct networks with properties similar to real systems. In order to construct WS network, first, a ring graph with n nodes is considered. Each node is connected to their k-nearest neighbors. Then, the links are rewired with probability p. One has the original ring graph for p = 0, while p = 1 results in a pure random graph. For the values of the rewiring probability between these two extremes, one may obtain a graph with both high clustering and short average path length. Such a property has been identified to be indeed the case in many real networks, especially social networks. We use the original preferential attachment algorithm proposed by Barabasi and Albert to construct BA networks, as follows. First, an all-to-all connected graph with m nodes is considered. Then, as each step, a new node is added to the network and m undirected links are created between this node and those already existing in the network. The probability of connecting the newly added nodes to an old node is proportional to the degree of the old node, the higher is the degree of an old node, the higher the probability of being tipped to a new node.

We first construct a network using either BA or WS models in layer $L_1$. Then, we copy all nodes of this network and a portion of its edges into layer $L_2$. The probability of existence of a link between two nodes (that are adjacent in $L_1$) in $L_2$ is set to PCP. Our aim is to predict the missing links in $L_2$, given the full details of $L_1$. This allows up testing whether the proposed framework is capable of correctly predicting the missing links. We consider BA networks with n=1000, m=5, and WS networks with n=1000, k=2, p=0.1.

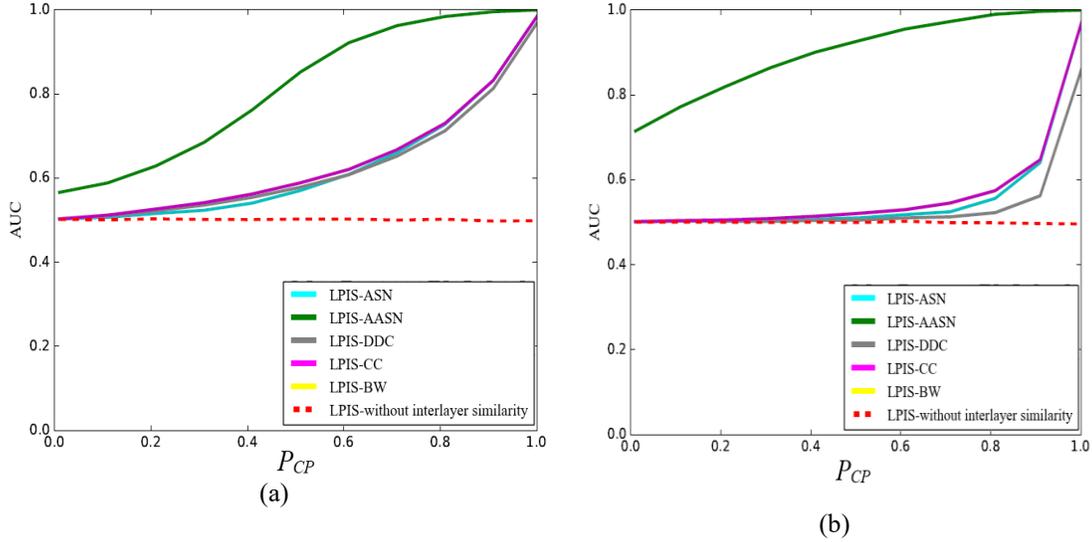

Fig 3. Impact of interlayer similarities on link prediction using the proposed framework LPIS with five interlayer similarity metrics; BW: Betweenness, CC: Clustering Coefficient, DDC: Degree-degree Correlation, ASN: Average Similarity of the Neighbors, and AASN: Asymmetric ASN. The network in the first layer is a) BA and b) WS; see text for detailed description of the way the multiplex networks are constructed.

Fig. 3 shows the results for both BA and WS networks. Five interlayer similarity measures are used in equation (16) to predict links in layer $L_2$. We also consider the case where no interlayer similarity is used, i.e. the predictions are performed based on only intralayer features. As it is seen, introducing the interlayer similarities to the prediction model always improves the link prediction performance as compared to the case with only intralayer features used for prediction. As expected, by increasing PCP, the interlayer similarity level increases, and thus the model performs with higher accuracy. Another observation is that the proposed interlayer similarity metric (AASN) significantly outperforms others by providing higher AUC in all cases. Indeed by taking into account the asymmetric effects in the definition of interlayer similarity works better than conventional symmetric metrics. In other words, the influence of layer $L_1$ to predict missing links in layer $L_2$ is not necessarily the same as the influence of layer $L_2$ to predict missing links in layer $L_1$. The proposed asymmetric similarity measure performs based on this idea. We next apply the proposed framework to predict missing links in real multiplex networks.

*4.2 Real Multiplex Networks*

In order to assess the performance of the proposed link prediction framework, we apply it to 4 real multiplex networks. These include a two-layer network of Twitter-Foursquare, two-layer network of Twitter-Instagram, 5-layer network of online and offline communications and the European air transportation network with 37 different layers. The Demographic information of the networks is shown in Table 1.

Table1. Demographic information of 4 real multiplex networks used in this work. Different layers of these networks are all scale-free with power-law degree distribution of $P(k) = k^{-\gamma}$, where γ is the power-law exponent. In order to obtain the value of this exponent, we fist a line to the corresponding intralayer degree distribution.

| Multiplex network | Layer | #nodes | #edges | Power-law exponent (γ) | Ratio of Common links | Link status |
|---|---|---|---|---|---|---|
| Twitter-Foursquare | Twitter | 1,565 | 2,663 | 3.58 | 0.55 | Directed |
|  | Foursquare | 1,565 | 36,056 | 2.01 | 0.098 | Undirected |
| Twitter-Instagram | Twitter | 13,298 | 52,668 | 3.69 | 0.45 | Directed |
|  | Instagram | 13,298 | 227,794 | 2.48 | 0.06 | Directed |
| Online and Offline Relationships | Facebook | 62 | 193 | 3.46 | 0.031 | Undirected |
|  | Leisure | 62 | 124 | 3.62 | 0.030 | Undirected |
|  | Work | 62 | 21 | 5.11 | 0.065 | Undirected |
|  | CO-authorship | 62 | 88 | 10.65 | 0.044 | Undirected |
|  | Launch | 62 | 194 | 4.56 | 0.031 | Undirected |
| EU-Air Transportation | Lufthansa | 450 | 244 | 2.79 | 0.015 | Undirected |
|  | Ryan air | 450 | 601 | 3.33 | 0.004 | Undirected |
|  | Easy jet | 450 | 307 | 2.50 | 0.011 | Undirected |
|  | British-airways | 450 | 66 | 7.26 | 0.016 | Undirected |
|  | Turkish-Airlines | 450 | 118 | 2.27 | 0.006 | Undirected |
|  | Air-Berlin | 450 | 184 | 2.01 | 0.013 | Undirected |
|  | Air-France | 450 | 69 | 2.64 | 0.018 | Undirected |
|  | Scandinavian-Airlines | 450 | 110 | 1.98 | 0.019 | Undirected |
|  | KLM | 450 | 62 | 2.74 | 0.019 | Undirected |
|  | Alitalia | 450 | 93 | 2.91 | 0.015 | Undirected |
|  | Swiss-International-Air-Lines | 450 | 60 | 3.89 | 0.015 | Undirected |
|  | Iberia | 450 | 35,450 | 4.44 | 0.031 | Undirected |
|  | Norwegian-Air-Shuttle | 450 | 87,450 | 2.23 | 0.017 | Undirected |
|  | Austrian-Airlines | 450 | 72,450 | 5.51 | 0.014 | Undirected |
|  | Flybe | 450 | 99,450 | 2.41 | 0.003 | Undirected |
|  | Wizz-Air | 450 | 92,450 | 2.57 | 0.006 | Undirected |
|  | TAP-Portougal | 450 | 53,450 | 6.46 | 0.013 | Undirected |
|  | Brussels-AIRLINES | 450 | 43,450 | 2.94 | 0.018 | Undirected |
|  | Finn air | 450 | 42,450 | 6.62 | 0.012 | Undirected |
|  | LOT-polish-Airlines | 450 | 55,450 | 2.91 | 0.014 | Undirected |
|  | Vueling-Airlines | 450 | 63,450 | 3.69 | 0.026 | Undirected |
|  | Air-nostrum | 450 | 69 | 2.58 | 0.017 | Undirected |
|  | Air-Lingus | 450 | 58,450 | 3.62 | 0.011 | Undirected |
|  | Germanwings | 450 | 67,450 | 3.03 | 0.009 | Undirected |
|  | Panagra-Airways | 450 | 58,450 | 2.65 | 0.014 | Undirected |
|  | Netjets | 450 | 180,450 | 2.80 | 0.007 | Undirected |
|  | Transavia-Holland | 450 | 57,450 | 3.43 | 0.014 | Undirected |
|  | Niki | 450 | 37,450 | 3.79 | 0.020 | Undirected |
|  | Sun express | 450 | 67,450 | 1.75 | 0.012 | Undirected |
|  | Aegean-Airlines | 450 | 53,450 | 2.35 | 0.019 | Undirected |
|  | Czech-Airlines | 450 | 41,450 | 3.27 | 0.011 | Undirected |
|  | European-Air-Transport | 450 | 73,450 | 3.27 | 0.011 | Undirected |
|  | Malev-Hungarian-Airlines | 450 | 34,450 | 6.44 | 0.013 | Undirected |
|  | Air-Baltic | 450 | 45,450 | 3.60 | 0.004 | Undirected |
|  | Wide roe | 450 | 90,450 | 2.44 | 0.001 | Undirected |
|  | TNT-Airways | 450 | 61,450 | 3.57 | 0.005 | Undirected |
|  | Olympic-Air | 450 | 43,450 | 2.32 | 0.015 | Undirected |

Twitter-Foursquare (TF) is a two-layer network from Twitter as a microblogging service and Foursquare as a location-based social network. This network includes 1565 users. The users on Twitter can share the tweets and also follow them, and thus this network is a directed one. In Foursquare network, the users can check in at any time, and share their views across different places. The communications of individuals in this network are undirected. This network has been used by Jalili et al. [22]. Twitter-Instagram (TI) is a two-layer social network including 3298 common nodes of Twitter and Instagram [41]. Instagram is a photo and video collaboration community that allows its users to upload their photos and videos to other social networks such as Facebook, Twitter and Flickr. The multiplex network of Online and Offline Relationships (OOR) considered here contains 5 layers Facebook, Leisure, Work, Co-authorship and Lunch among employees of the Department of Computer Science in Aarhus University [42]. The offline data of this network has been gathered through questionnaires distributed online to 62 employees. The EU air transportation network consists 37 different layers, one for each airline [43]. Each of these layers includes 450 nodes, representing the airports, and the connections at each of these layers are the inter-city routs for that airline.

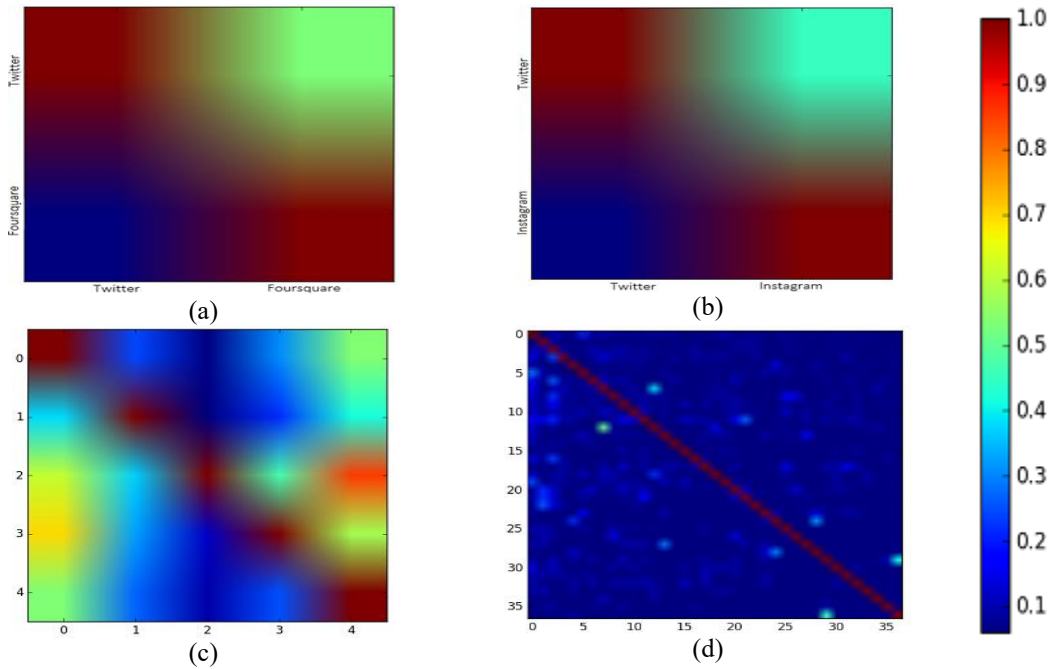

Fig 4. Visualizing the Asymmetric ASN values from each pair of layers for the real multiplex networks considered in this work; (a) Twitter-Foursquare, (b) Twitter-Instagram, (c), Online-Offline Relationship and (d) EU-Air Transportation.

Our proposed framework is based on interlayer similarity values. Fig. 4 shows the asymmetric ASN, which is proposed to use for the link prediction task. Tables 2-5 show the link prediction results using the proposed framework LPIS. Various interlayer similarity metrics are considered including Betweenness (BW), Clustering Coefficient (CC), Degree-degree Correlation (DDC), Average Similarity of the Neighbors (ASN) and Asymmetric ASN (AASN). Furthermore, three different intralayer features (Adamic-Adar, Jaccard Coefficient and Preferential Attachment) are considered in the prediction process. The performance of LPIS is considered with two state-of-the-art methods recently proposed for the link prediction in multiplex networks. These include Entropy Based Method (EBM) [23] and Meta Path Based Method (MPBM) [23].

12Table 2. AUC values for the multiplex networks considered in this work including Twitter-Foursquare (TF), Twitter-Instagram (TI), and Online and Offline Relationship (OOR). TF – Twitter means that the link prediction is studied for Twitter layer of TF given the intralayer features and interlayer similarity. The proposed method, Link Prediction accounting Interlayer Similarity (LPIS), with various interlayer similarity metrics (BW: BetWeenness, CC: Clustering Coefficient, DDC: Degree-Degree Correlation, ASN: Average Similarity of the Neighbors, and AASN: Asymmetric ASN) is applied to the datasets and compared with two state-of-the-art methods recently proposed for the link prediction task of multiplex networks. The intralayer feature is considered as Adamic-Adar index. The best algorithm is highlighted in bold font for each case.

| Multiplex network-Layer | LPIS with interlayer similarity as: | | | | | EBM [24] | MPBM [23] |
|---|---|---|---|---|---|---|---|
| | BW | CC | DDC | ASN | AASN | | |
| TF – Twitter | **0.896** | 0.72 | 0.266 | 0.25 | 0.884 | 0.842 | 0.773 |
| TF – Foursquare | **0.945** | 0.928 | 0.925 | 0.936 | 0.938 | 0.864 | 0.934 |
| TI – Twitter | 0.986 | 0.917 | 0.341 | 0.61 | **0.995** | 0.934 | 0.813 |
| TI – Instagram | 0.955 | 0.929 | 0.936 | 0.964 | 0.964 | 0.904 | 0.939 |
| OOR – Facebook | **0.953** | 0.874 | 0.90 | 0.868 | 0.898 | 0.840 | 0.927 |
| OOR – Leisure | **0.924** | 0.833 | 0.795 | 0.855 | 0.898 | 0.853 | 0.882 |
| OOR – Work | 0.967 | 0.842 | 0.233 | 0.257 | **0.968** | 0.781 | 0.603 |
| OOR – Coauthorship | **0.952** | 0.79 | 0.759 | 0.653 | 0.941 | 0.82 | 0.739 |
| OOR – Lunch | **0.842** | 0.772 | 0.716 | `0.70 | 0.789 | 0.761 | 0.832 |

Table 3. AUC values for the prediction algorithms with Jaccard Coefficient as the intralayer feature. Other designations are as Table 2.

| Multiplex network-Layer | LPIS with interlayer similarity as: | | | | | EBM [24] | MPBM [23] |
|---|---|---|---|---|---|---|---|
| | BW | CC | DDC | ASN | AASN | | |
| TF – Twitter | **0.887** | 0.72 | 0.33 | 0.25 | 0.885 | 0.84 | 0.773 |
| TF – Foursquare | **0.929** | 0.91 | 0.908 | 0.924 | 0.925 | 0.915 | 0.919 |
| TI – Twitter | 0.987 | 0.955 | 0.437 | 0.705 | **0.998** | 0.949 | 0.812 |
| TI – Instagram | 0.884 | 0.809 | 0.81 | 0.899 | 0.899 | 0.867 | 0.875 |
| OOR – Facebook | **0.953** | 0.879 | 0.901 | 0.873 | 0.901 | 0.847 | 0.93 |
| OOR – Leisure | **0.926** | 0.890 | 0.799 | 0.863 | 0.902 | 0.838 | 0.879 |
| OOR – Work | 0.966 | 0.868 | 0.227 | 0.254 | **0.974** | 0.769 | 0.601 |
| OOR – Coauthorship | **0.956** | 0.860 | 0.805 | 0.795 | 0.952 | 0.832 | 0.743 |
| OOR – Lunch | **0.835** | 0.765 | 0.706 | 0.692 | 0.784 | 0.751 | 0.833 |

Table 4. AUC values for the prediction algorithms with Preferential Attachment as the intralayer feature. Other designations are as Table 2.

| Multiplex network-Layer | LPIS with interlayer similarity as: | | | | | EBM [24] | MPBM [23] |
|---|---|---|---|---|---|---|---|
| | BW | CC | DDC | ASN | AASN | | |
| TF – Twitter | **0.933** | 0.767 | 0.25 | 0.345 | 0.881 | 0.736 | 0.877 |
| TF – Foursquare | **0.918** | 0.899 | 0.883 | 0.894 | 0.897 | 0.778 | 0.916 |
| TI – Twitter | **0.994** | 0.901 | 0.306 | 0.537 | 0.991 | 0.583 | 0.941 |
| TI – Instagram | **0.942** | 0.912 | 0.889 | 0.938 | 0.938 | 0.695 | 0.939 |
| OOR – Facebook | 0.647 | 0.555 | 0.558 | 0.527 | **0.74** | 0.508 | 0.734 |
| OOR – Leisure | **0.914** | 0.911 | 0.888 | 0.891 | 0.903 | 0.628 | 0.908 |
| OOR – Work | 0.889 | 0.716 | 0.194 | 0.249 | **0.91** | 0.369 | 0.378 |
| OOR – Coauthorship | **0.785** | 0.69 | 0.663 | 0.644 | 0.759 | 0.667 | 0.706 |
| OOR – Lunch | **0.772** | 0.688 | 0.683 | 0.665 | 0.713 | 0.603 | 0.765 |

Table 5. AUC values for the prediction algorithms with Local Path as the intralayer feature. Other designations are as Table 2.

| Multiplex network-Layer | LPIS with interlayer similarity as: | | | | | EBM [24] | MPBM [23] |
|---|---|---|---|---|---|---|---|
| | BW | CC | DDC | ASN | AASN | | |
| TF – Twitter | **0.863** | 0.698 | 0.411 | 0.328 | 0774 | 0.813 | 0.816 |
| TF – Foursquare | 0.845 | 0.846 | 0.845 | 0.838 | 0.842 | 0.841 | **0.849** |
| TI – Twitter | **0.883** | 0.813 | 0.167 | 0.246 | 0.821 | 0.868 | 0.793 |
| TI – Instagram | **0.897** | 0.845 | 0.862 | 0.854 | 0.854 | 0.858 | 0.859 |
| OOR – Facebook | **0.747** | 0.743 | 0.745 | 0.746 | 0.745 | 0.733 | 0.764 |
| OOR – Leisure | 0.773 | 0.773 | 0.767 | 0.756 | **0.776** | 0.758 | 0.774 |
| OOR – Work | 0.731 | 0.449 | 0.461 | 0.354 | 0.694 | 0.727 | **0.75** |
| OOR – Coauthorship | **0.764** | 0.76 | 0.761 | .762 | 0.763 | 0.744 | 0.761 |
| OOR – Lunch | **0.764** | 0.751 | 0.724 | 0.713 | 0.752 | 0.752 | 0.762 |

Table 6. AUC values for the prediction algorithms with Resource Allocation as the intralayer feature. Other designations are as Table 2.

| Multiplex network-Layer | LPIS with interlayer similarity as: | | | | | EBM [24] | MPBM [23] |
|---|---|---|---|---|---|---|---|
| | BW | CC | DDC | ASN | AASN | | |
| TF – Twitter | 0.696 | 0.634 | 0.314 | 0.28 | 0.658 | 0.778 | **0.781** |
| TF – Foursquare | 0.842 | 0.848 | 0.866 | 0.859 | 0.858 | 0.869 | **0.873** |
| TI – Twitter | **0.795** | 0.710 | 0.213 | 0.287 | 0.730 | 0.792 | 0.793 |
| TI – Instagram | **0.88** | 0.853 | 0.852 | 0.85 | 0.876 | 0.853 | 0.872 |
| OOR – Facebook | 0.795 | 0.816 | 0.808 | **0.824** | 0.817 | 0.715 | 0.821 |
| OOR – Leisure | 0.789 | 0.804 | 0.822 | 0.836 | **0.838** | 0.796 | 0.799 |
| OOR – Work | 0.568 | 0.684 | 0.383 | 0.548 | **0.877** | 0.634 | 0.488 |
| OOR – Coauthorship | 0.719 | 0.716 | 0.714 | 0.804 | **0.807** | 0.734 | 0.709 |
| OOR – Lunch | 0.572 | 0.585 | 0.658 | 0.659 | 0.711 | 0.730 | **0.793** |

Tables 2-6 show the link prediction results for TF, TI and OOR networks with intralayer feature selected as Adamic-Adar, Jaccard Coefficient, Preferential Attachment, Local Path and Resource Allocation respectively. As it is seen, LPIS with BW and AASN works betters than the case with other three interlayer similarity metrics used in the prediction framework. BW is a metric requiring global information of the networks, while AASN requires only local information on nodes, and as such is much simpler to compute than BW. This is especially important for large-scale networks, for which the local measures are the only choice in some cases. LPIS with BW or AASN significantly outperform the state-of-the-art methods. Table 7 shows the results for the EU air transportation network. Adamic-Adar is used as the intralayer feature and the results for other intralayer features are similar (data not shown here). Although the interlayer similarity values for this network are less than the other three networks. The proposed framework with AASN, ASN, and BW interlayer similarities significantly outperforms EBM and MPBM. These results reveal superiority of the proposed framework for link prediction in multiplex networks.

14141414141414

Table 7. AUC values for the EU-Air Transportation network with Adamic-Adar as the intralayer feature. Other designations are as Table 2.

| Layer of EU-Air Transportation | LPIS with interlayer similarity as: | | | | | EBM [23] | MPBM [22] |
|---|---|---|---|---|---|---|---|
| | BW | CC | DDC | ASN | AASN | | |
| Lufthansa | 0.963 | 0.924 | 0.592 | 0.980 | **0.982** | 0.914 | 0.816 |
| Ryan air | 0.961 | 0.951 | 0.805 | 0.966 | **0.966** | 0.826 | 0.837 |
| Easy jet | **0.981** | 0.960 | 0.726 | 0.973 | 0.976 | 0.853 | 0.878 |
| British-airways | 0.992 | 0.419 | 0.242 | 0.994 | **0.995** | 0.889 | 0.608 |
| Turkish-Airlines | 0.871 | 0.963 | 0.369 | 0.985 | **0.987** | 0.656 | 0.592 |
| Air-Berlin | 0.982 | 0.926 | 0.783 | 0.986 | **0.988** | 0.883 | 0.810 |
| Air-France | 0.933 | 0.896 | 0.083 | 0.987 | **0.991** | 0.854 | 0.712 |
| Scandinavian-Airlines | 0.987 | 0.920 | 0.694 | 0.989 | **0.992** | 0.852 | 0.915 |
| KLM | 0.993 | 0.919 | 0.197 | 0.995 | **0.996** | 0.915 | 0.556 |
| Alitalia | 0.991 | 0.857 | 0.594 | 0.993 | **0.994** | 0.969 | 0.749 |
| Swiss-International-Air-Lines | 0.987 | 0.913 | 0.532 | 0.991 | **0.994** | 0.977 | 0.720 |
| Iberia | 0.882 | 0.944 | 0.218 | 0.985 | **0.991** | 0.908 | 0.445 |
| Norwegian-Air-Shuttle | 0.937 | 0.856 | 0.819 | 0.986 | **0.988** | 0.757 | 0.798 |
| Austrian-Airlines | 0.979 | 0.939 | 0.653 | 0.993 | **0.994** | 0.905 | 0.818 |
| Flybe | 0.977 | 0.691 | 0.673 | 0.990 | **0.991** | 0.810 | 0.721 |
| Wizz-Air | 0.942 | 0.93 | 0.651 | 0.971 | **0.973** | 0.491 | 0.548 |
| TAP-Portugal | 0.987 | 0.930 | 0.761 | 0.992 | **0.995** | 0.794 | 0.583 |
| Brussels-AIRLINES | 0.989 | 0.983 | 0.216 | 0.991 | **0.996** | 0.959 | 0.900 |
| Finn air | 0.96 | 0.792 | 0.381 | 0.994 | **0.997** | 0.818 | 0.616 |
| LOT-polish-Airlines | 0.984 | 0.864 | 0.442 | 0.992 | **0.995** | 0.917 | 0.555 |
| Vueling-Airlines | 0.990 | 0.975 | 0.643 | 0.992 | **0.993** | 0.917 | 0.849 |
| Air-nostrum | 0.980 | 0.930 | 0.316 | 0.987 | **0.991** | 0.760 | 0.654 |
| Air-Lingus | 0.986 | 0.897 | 0.578 | 0.980 | **0.990** | 0.906 | 631 |
| Germanwings | 0.975 | 0.909 | 0.692 | 0.991 | **0.994** | 0.938 | 0845 |
| Panagra-Airways | 0.984 | 0.827 | 0.442 | 0.992 | **0.995** | 0.737 | 0.582 |
| Netjets | 0.928 | 0.853 | 0.756 | 0.953 | **0.956** | 0.750 | 0.692 |
| Transavia-Holland | 0.926 | 0.844 | 0.636 | 0.990 | **0.992** | 0.749 | 0.749 |
| Niki | 0.988 | 0.910 | 0.631 | 0.993 | **0.996** | 0.892 | 0.583 |
| Sun express | 0.982 | 0.887 | 0.404 | 0.992 | **0.994** | 0.747 | 0.883 |
| Aegean-Airlines | 0.930 | 0.847 | 0.413 | 0.991 | **0.993** | 0.825 | 0.733 |
| Czech-Airlines | 0.988 | 0.913 | 0.215 | 0.995 | **0.996** | 0.908 | 0.778 |
| European-Air- Transport | 0.930 | 0.882 | 0.405 | 0.982 | **0.985** | 0.825 | 0.600 |
| Malev-Hungarian-Airlines | 0.987 | 0.985 | 0.143 | 0.993 | **0.997** | 0.975 | 0.546 |
| Air- Baltic | 0.930 | 0.757 | 0.459 | 0.992 | **0.993** | 0.775 | 0.530 |
| Wide roe | 0.922 | 0.892 | 0.902 | 0.975 | **0.987** | 0.764 | 0.781 |
| TNT-Airways | 0.867 | 0.814 | 0.523 | 0.978 | **0.981** | 0.631 | 0.545 |
| Olympic-Air | 0.898 | 0.657 | 0.238 | 0.991 | **0.996** | 0.559 | 0.538 |

## 5. Conclusions

Many real networks develop connections between individual units in multiple layers and can be better modeled by multilayer (or multiplex) networks. In this paper, we proposed a novel framework for predicting missing links in multiplex networks. The question is how to predict the missing links in one of the layers of a multiplex network taking into account the information available from other layers. Real multiplex networks often show a significant interlayer similarity of connections. The proposed link prediction method introduces a systematic approach to take into account interlayer similarities in the link prediction process. It is composed of two parts: intralayer features and interlayer similarity. The intralayer features are proximity-based features such as Adamic-Adar or Jaccard Coefficient that are obtained based on the structural information of the layer for which the link prediction is performed. These features are then properly combined with interlayer similarity to effectively use information of other layers in the link prediction process. We applied the proposed framework on both synthetic and real multiplex networks. The experimental results showed that the proposed link prediction framework outperforms state-of-the-art methods.

6.  Acknowledgements

This work was supported in part by a grant from IPM (No.CS1396-4-49) and Mahdi Jalili is supported by Australian Research Council through project No DP170102303.